# Transfer Characteristics in Graphene Field-Effect Transistors with Co Contacts

Ryo Nouchi[1,a], Masashi Shiraishi[1,2], and Yoshishige Suzuki[1]

[1]*Division of Materials Physics, Graduate School of Engineering Science, Osaka University, Machikaneyama-cho 1-3, Toyonaka 560-8531, Japan*

[2]*PRESTO, JST, 4-1-8 Honcho, Kawaguchi 332-0012, Saitama, Japan*

Graphene field-effect transistors with Co contacts as source and drain electrodes show anomalous distorted transfer characteristics. The anomaly appears only in short-channel devices (shorter than approximately 3 μm) and originates from a contact-induced effect. Band alteration of a graphene channel by the contacts is discussed as a possible mechanism for the anomalous characteristics observed.

[a] Author to whom correspondence should be addressed; present address: WPI-Advanced Institute for Materials Research, Tohoku University, Sendai 980-8577, Japan; electronic mail: nouchi@sspns.phys.tohoku.ac.jp

Graphene, one-atomic carbon sheet with a honeycomb structure, has been attracting significant attention due to its unique physical properties, such as the massless Dirac fermion system.[1] This material shows an extraordinarily high carrier mobility of more than 200,000 cm$^2$ V$^{-1}$ s$^{-1}$ (ref. 2), and is considered to be a major candidate for a future high-speed transistor material. In addition, graphene has shown its ability to transport charge carriers with spin coherence even at room temperature,[3-5] and is regarded as a pivotal material in the emerging field of molecular spin electronics. In order to construct such electronic devices, metallic materials should make a contact with the graphene layers. The effect of metal contacts can be detected using the structure of a field-effect transistor (FET) and measuring the transfer characteristics (drain current, $I_D$, vs. gate voltage, $V_G$, characteristics). For instance, the difference between the drain currents of graphene FETs at exactly opposite charge densities (at the same carrier densities with opposite charge polarities) has been explained by a metal-contact effect.[6] Charge transfer from metal to graphene leads to a *p-p*, *n-n* or *p-n* junction[7,8] in graphene, depending on the polarity of carriers in the bulk of the graphene sheet. An additional resistance arises as a result of the density step created along the graphene channel, which causes asymmetry.

In this letter, we report the effect of metal contacts on the transfer characteristics of graphene FETs. In particular, the choice of metal and the gap between the metal contacts (source and drain electrodes) have been examined by employing a FET structure. It was found that graphene FETs with Co contacts and short channels display anomalous distorted transfer characteristics, indicating that the anomaly originates from Co contacts.

Graphene layers were formed onto a highly-doped Si substrate with a 300 nm thick thermal oxide layer using conventional mechanical exfoliation.[9] The starting graphite crystal used was Super Graphite® from Kaneka Corporation.[10] The thicknesses of the graphene layers were determined to be approximately 1 nm by atomic force microscopic observations in tapping mode. These layers were determined to be one-atom thick from the optical contrast.[11] Metal electrodes (Co and Au) were fabricated onto the graphene layers by electron beam lithography and liftoff techniques. For the Au electrodes, 5 nm thick Cr was deposited as an adhesive layer prior to Au deposition. The electrodes fabricated in this study had a total thickness of 50 nm. The FET characteristics were measured in low vacuum at room temperature.

The graphene FET device structure is schematically displayed in Fig. 1(a), and the transfer characteristics are shown in Figs. 1(b) and 1(c) for Cr/Au and Co source/drain electrodes, respectively. Cr/Au is a conventionally used metallic material for electronic devices, and Co is a popular material for spin-electronic devices as a source of spin-polarized current. Although the graphene FET with Cr/Au contacts exhibits conventional transfer characteristics, as widely reported previously, that with the Co contacts displays anomalous distorted characteristics, especially in the negatively gated region. The distortion disappeared when the gap between the metal contacts (channel length) was lengthened (Fig. 2), which indicates a contact-related effect. The shorter channel results in lower channel resistance, and the resistance originating from the contacts should have a more dominant effect on the two-terminal resistance. In fact, the resistances at the $I_D$ minima are not proportional to the ratio of channel length to channel width, and thus the contact-related effects contribute to the device resistance.

Distorted transfer characteristics can be observed in a double-gated device under a specific relation between the two gate voltages.[12] In such a device, the graphene channel is divided into two parts: a double-gated region and a bottom-gated region. By applying a gate voltage from the top gate electrode, the Fermi level of the top-gated region is shifted, compared to the rest of the channel. The resulting transfer characteristics display two local minima when the shift is sufficiently large. If such a large shift occurs due to the charge transfer from the metal electrodes, devices without a top gate also appear to display transfer characteristics with two local minima. However, this is not the case for the Co-contacted devices. In the wide-ranging transfer characteristics of the 3 μm channel device (Fig. 3), decreases in the current compared with ordinary transfer characteristics can be seen at gate voltages of -70 and +30 V in addition to the minimum at a gate voltage of +2 V. The minimum at +2 V is considered to be the charge neutrality point (the so-called Dirac point); the other two anomalous points are therefore a consequence of the metal contacts. Such additional decreases can also be distinguished in the 2 μm channel device [Fig. 1(c)], although the appearance of the anomaly around 30 V is somewhat weaker than that of the 3 μm channel device. The difference in the strength of the anomaly might be caused by contamination of the graphene surface, which would vary from device to device and cause the difference in the strength of the Co-graphene interaction.

Generally accepted values of work functions of polycrystalline Au and Co are 5.1 and 5.0 eV, respectively.[13] These values are higher than that of graphene [4.6 eV (ref. 14)] and there should exist an electron transfer from graphene to the metals. However, the two metallic materials have very similar work functions; thus the simple ionic charge transfer effect alone cannot account for the observed difference in the shape of transfer characteristics. A recent first-principle calculation at the level of density functional theory[15] indicates a strong chemical interaction between Co and graphene, where conical points at $K$ disappear, and instead a mixed metal-graphene character appears in the electronic structure. From examination of Fig. 3, two peaks centered at -40 and +20 V can be distinguished. The Fermi energy shifts from the Dirac point at +2 V are determined to be -0.20 and +0.13 eV, respectively, using the relationship $\Delta E_F = \text{sgn}(n)\hbar v_F \sqrt{\pi |n|}$, where $\hbar$ is the Dirac constant, $v_F$ is the Fermi velocity (around $10^6$ m s$^{-1}$), and $n$ is the two-dimensional charge density.[16,17] The gate voltage changes $n$ through the relation $n = \varepsilon_r \varepsilon_0 (V_G - V_{G0})/d$, where $\varepsilon_r$ is the relative permittivity of the gate dielectric (3.8 for SiO$_2$), $\varepsilon_0$ is the electric constant, $V_{G0}$ is the gate voltage corresponding to the Dirac point, and $d$ is the thickness of the gate dielectric (300 nm in this study). Similar peaks have been reported in scanning tunneling spectroscopic data of a graphene layer formed on Ru(0001), where two peaks were observed at -0.4 and +0.2 V.[18] The peaks were not observed with the bare Ru surface, and were therefore considered to originate from the metal contact. A band structure for the graphene/Ni(111) system has been calculated using a discrete variational X$\alpha$ method, and a gap of approximately 1.0 eV was expected to open at the $K$ points.[19] Such a gap opening affects the transfer characteristics and should be detected as an increase in $I_D$ at gate voltages corresponding to the edges of the gap. However, the transfer characteristics in Fig. 3 display a decrease in $I_D$ around -70 V, which cannot be explained by a simple gap opening. This decrease is possibly explained by contact-induced states formed by the hybridization of graphene π and Co *d* bands, which may form some resonant states and cause the negative $dI_D / d|V_G - V_{G0}|$ values.

Another possible mechanism is the diffusion of Co atoms into/onto graphene channels. In a single charge tunneling device of a single CdTe nanorod with Cr/Pd contacts, a chemical transformation was found to occur by the diffusion of Pd atoms 20-30 nm into the nanorod.[20] However, the robust honeycomb lattice structure of graphene and the possibly strong chemical interactions at Co/graphene interfaces should prevent Co atoms from diffusing a long distance into and onto graphene channels.

In summary, the effect of metallic electrode materials contacting graphene channel layers was studied using FET structures. Cr/Au and Co contacts were investigated, and it was found that graphene FETs with Co contacts and short channels exhibit distorted transfer characteristics that have two peaks at -0.20 and +0.13 eV, in addition to the common minimum at the Dirac point. The anomalous distortion can be considered to be the result of band alteration of the graphene channel underneath the Co contacts. The present study ascertained the metal-induced alteration of the FET characteristics of graphene. These results indicate particularly crucial issues for the development of future graphene microelectronics that consist of short-channel devices.

We acknowledge Dr. Mutsuaki Murakami of Kaneka Corporation for providing the Super Graphite® Crystal. This work was supported by a grant from the Foundation Advanced Technology Institute.

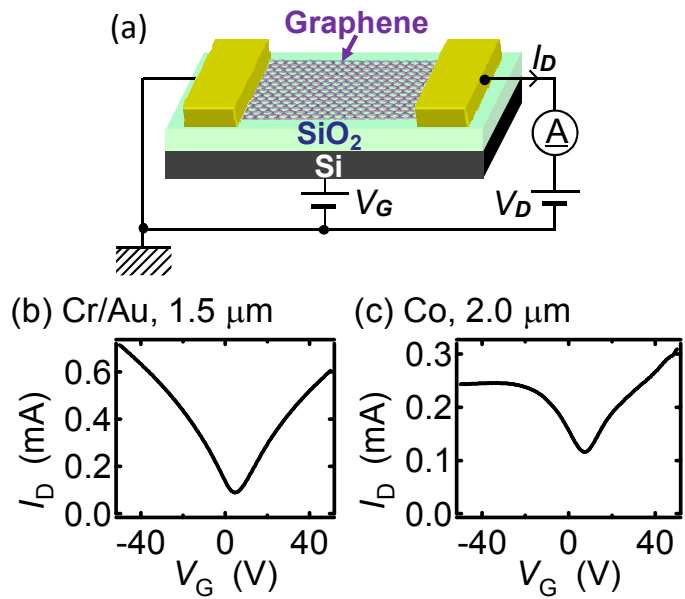

FIG. 1. (Color online) (a) Schematic diagram of a graphene FET. (b) Transfer characteristics of a graphene FET with Cr/Au electrodes. The channel length was 1.5 μm. (c) Transfer characteristics of a graphene FET with Co electrodes. The channel length was 2.0 μm. An anomalous distortion is clearly seen in the characteristics.

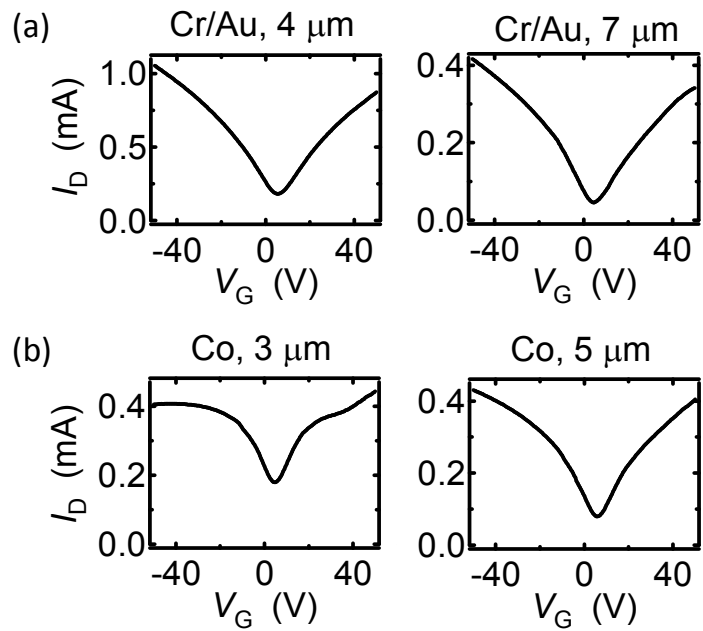

FIG. 2. Transfer characteristics of longer-channel graphene FETs with (a) Cr/Au and (b) Co contacts. The distortion is barely observed in the 5 μm channel device, even with Co contacts.

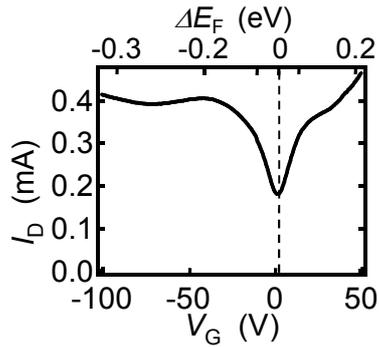

FIG. 3. Wide-ranging transfer characteristics of Co-contacted graphene FETs with a channel length of 3 μm. The device is identical to that shown in the left panel of Fig. 2(b). In order to avoid an accidental breakdown of the device, the upper limit of the gate voltage was set to 50 V. The gate voltage corresponding to the charge neutrality was determined to be identical to that showing the smallest current.